# Origin of perpendicular magnetic anisotropy in ultra-thin metal films studied by *in-situ* neutron reflectometry

Grigorii Kirichuk[a], Alexey Grunin[a], Artur Dolgoborodov[b], Pavel Prokopovich[a], Petr Shvets[a], Alexey Vorobiev[c], Anton Devishvilli[d], Alexandr Goikhman[b], Ksenia Maksimova[b]

[a]Research and Educational Center "Functional Nanomaterials", Immanuel Kant Baltic Federal University, 236041, Kaliningrad, Russian Federation

[b]Koenigssystems UG, 22869, Schenefeld bei Hamburg, Germany.

[c]NanoLund and Physical Chemistry, Lund University, PO Box 124, 22 100Lund, Sweden

[d]Institut Laue-Langevin, 71 avenue des Martyrs, 38000 Grenoble, France

*Abstract*

Perpendicular magnetic anisotropy (PMA) plays an important role in different spintronic devices. The rapid development of spintronics requires a better understanding of the nature and mechanisms of the PMA formation. In our article, we demonstrate the potential of studying PMA by in-situ polarized neutron reflectometry combined with pulsed laser deposition. Using these techniques, we show the formation of out-of-plane anisotropy in thin CoFeB films (11–18 Å) with a capping Mo layer. Investigating thick (53 Å) and thin (5 Å) molybdenum films, we demonstrate that in both cases, PMA was established in bilayer structures without any thermal annealing. Also, we demonstrate how an additional silicon layer grown over the CoFeB / Mo bilayer can critically alter the magnetic properties of the sample. Such studies are possible only through the unique combination of the growth method with the in-situ polarized neutron reflectometry measurement technique. We believe that this approach will open up broad opportunities for the development and investigation of new spintronic devices.

*Introduction*

The structures with perpendicular magnetic anisotropy (PMA) are widely used, but the mechanisms of PMA formation are still under discussion [1-3]. PMA appears in ultrathin films or their multilayer structures, which makes the investigation of this phenomenon rather difficult [4-6]. The most common materials to create PMA are combinations based on MgO / CoFeB [7,8]. Various capping layers of heavy metals (HM) such as Mo [9,10], Hf [11,12], W [13,14], Ta [7,15], and Pt [16,17] have been used to improve the magnetic properties. More complex structures like CoFeB / MgO / CoFeB [18,19], MgO / CoFeB / HM / CoFeB / MgO [20], SiO$_2$ /

W / CoFeB / W [21], and $Co_{75}Mn_{25}$ / Mo / CoFeB multilayers [22] have also been investigated. In such structures, the formation of PMA significantly depends on the ferromagnet / oxide (FM / Ox) interface [7]. At the same time, the ferromagnet / heavy metal (FM / HM) interface might be equally important for magnetic properties [15]. Moreover, this FM / HM interface is essential for the appearance of the spin-orbit torque (SOT) effect, which is the fundamental basis of the spin Hall and the anomalous spin Hall effects [6, 23, 24]. The investigation and application of SOT have become a new stage in the development of spintronics based on metallic structures [23-26]. Therefore, it is crucial to study the mechanism of the formation of perpendicular magnetic anisotropy in metallic bilayer FM / HM structures.

In this paper, we demonstrate a layer-by-layer growth and in-situ study of the evolution of magnetic properties in CoFeB / Mo films using polarized neutron reflectometry (PNR). Thin films were produced by the pulsed laser deposition (PLD) in a specially designed vacuum chamber, which was placed in the gap between the magnets at the PNR measurement station SuperADAM, Institut Laue-Langevin (ILL), Grenoble, France. The combination of PLD and PNR allowed us to grow ultrathin films with high accuracy and simultaneously characterize their magnetic properties in the same vacuum chamber [14, 27]. First, we studied single-layer CoFeB thin films with thicknesses approximately of 18 Å. Then, we expanded our structures by growing additional layers and re-examined them by PNR. In our work, we demonstrated the formation of PMA in CoFeB-based structures with a Mo capping layer of varying thickness. For such bilayer structures, we observed PMA when the molybdenum layer was as thin as just 5 Å. The discovery of PMA in such ultra-thin metal films was possible only thanks to the possibility of characterizing the samples by in-situ neutron reflectometry directly in the PLD growth chamber.

*Experimental details*

Samples were produced by PLD and characterized by PNR at the SuperADAM station, Institut Laue-Langevin (ILL), Grenoble, France [28]. We designed a vacuum chamber for the deposition and placed it between the magnets of the measurement line of the station (Figure 1). Thus, we were able to grow and investigate the sample at the same time in the same vacuum volume.

A solid-state Nd:YAG laser was used for deposition, equipped with additional optics to generate the fourth harmonic (266 nm). The pulse repetition frequency was 10 Hz, with each pulse having an energy of 30 mJ and a duration of 7 ns. The beam was focused to a spot with a diameter of 1 mm. The sample rotation speed was 10 rpm, and the distance between the target and the sample was 7 cm. The CoFeB, Mo, and Si thin films were obtained from the corresponding stoichiometric targets $Co_{40}Fe_{40}B_{20}$ (target purity 99.99%), molybdenum (target purity 99.99%), and silicon (target purity 99.99%). The films were grown on a (100)-oriented single-crystal silicon with thermal oxide $SiO_2$ with a thickness of ~ 750 nm. The size of the silicon substrate was 20x20 mm$^2$. The deposition was carried out in a high vacuum with a base pressure of $1 \times 10^{-8}$ mBar and at room temperature. Neutron reflectivity measurements were conducted under the same conditions. The wavelength of the monochromatic neutron beam was 5.21 Å, with a polarization of 99.8%. Magnetic fields of 100 and 7500 Oe were used for the measurements. A detailed description of the PNR measurement station SuperADAM and the methodology for neutron reflectometry can be found elsewhere [28-30]. The fitting of the collected data was carried out using the open-source GenX software [31].

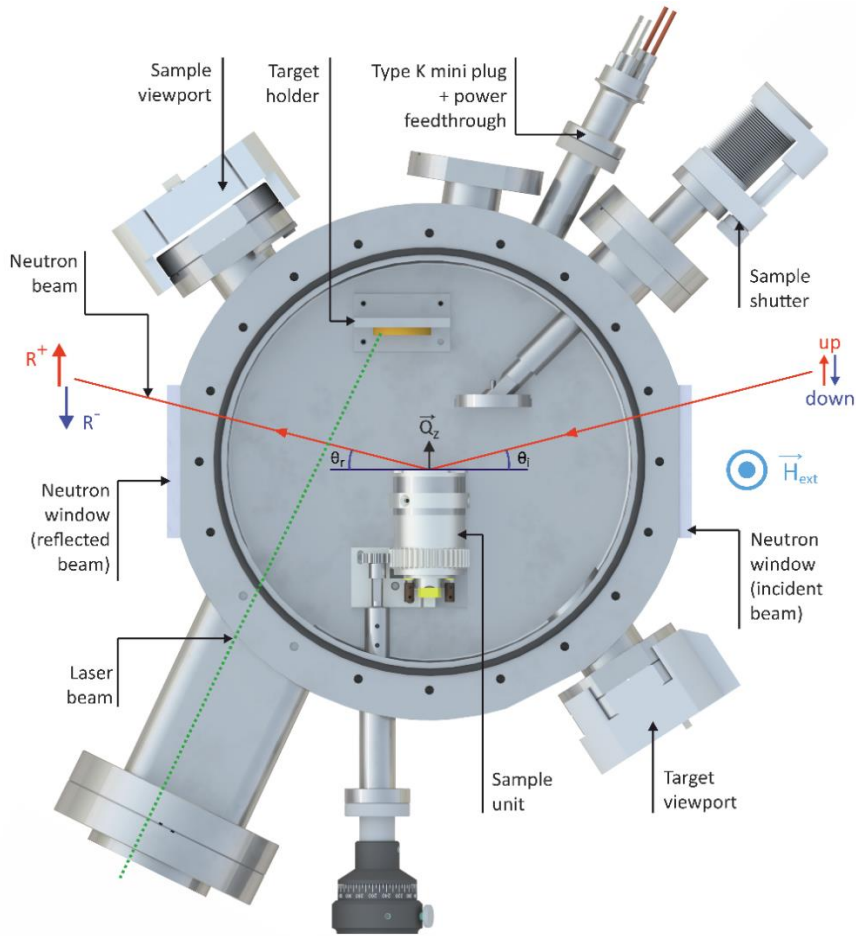

*Fig. 1. Schematics of the PLD setup for in-situ neutron reflectometry measurements. A target is sputtered by a laser beam (the green dotted line) forming a film on a substrate. A neutron beam (the red line) passes through the aluminum windows of the chamber and reflects from the sample yielding a reflectivity curve.*

*Results and discussion*

1. *PMA in CoFeB / Mo (thick HM layer)*

First, we produced a single-layer thin CoFeB film and characterized it by neutron reflectivity. The film thickness determined from the PNR measurements was $18 \pm 2$ Å. From neutron reflectometry curves measured at 100 Oe ($R^+$ and $R^-$ polarizations, Figure 2), we estimated the magnetic scattering length density (mSLD) as $(1.08 \pm 0.13) \times 10^{-6}$ Å$^{-2}$, corresponding to the magnetization of $380 \pm 40$ emu/cm$^3$ (1 emu/cm$^3$ = $10^3$ A/m) [32,33]. The nuclear SLD (nSLD) for the CoFeB film is $(5.71 \pm 0.15) \times 10^{-6}$ Å$^{-2}$. The inset in Figure 2 shows the spin asymmetry (SA) curve, calculated using the formula [34]:

$$SA = \frac{R^+ - R^-}{R^+ + R^-}$$

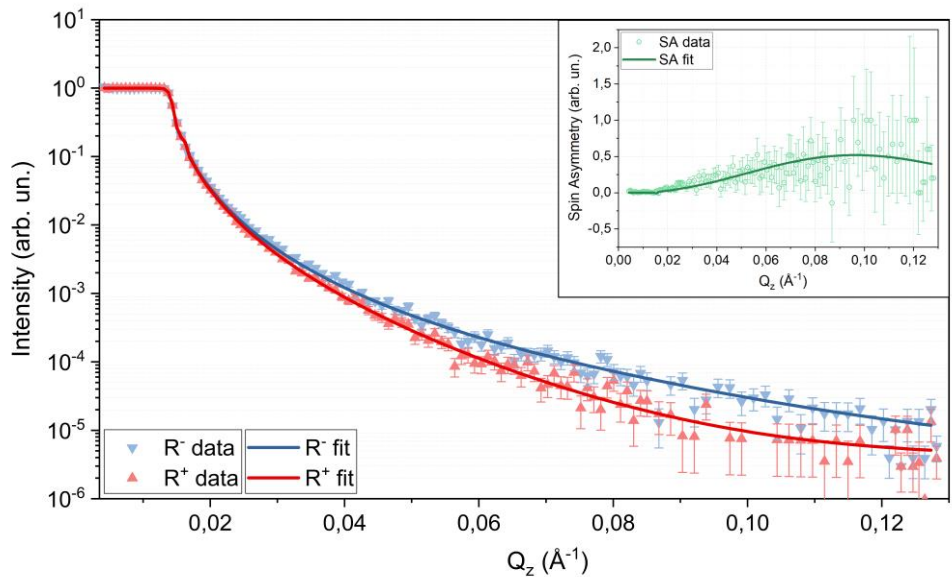

*Fig. 2. Neutron reflectivity curves for $R^+$ and $R^-$ polarizations for a CoFeB film, measured at a magnetic field of 100 Oe (the measured data are represented by dotted lines and the fitted models are represented by solid lines). The inset shows the spin asymmetry curve.*

After the measurements, we deposited a molybdenum film on the previously grown CoFeB film. From PNR curves for the bilayer structure (Figure 3a), we determined that the thickness of the Mo layer was $51 \pm 2$ Å. Compared to the single CoFeB film (Figure 2), the CoFeB/Mo bilayer structure did not show any splitting between the $R^+$ and $R^-$ curves at 100 Oe. However, the splitting became obvious when we increased the magnetic field to 7500 Oe. From the nSLD and mSLD profiles for the bilayer structure measured at a field of 7500 Oe (Figure 3b), we determined the value of mSLD as $(1.04 \pm 0.13) \times 10^{-6}$ Å$^{-2}$, corresponding to $360 \pm 30$

emu/cm$^3$. The nSLD for the molybdenum layer is equal to $(4.74 \pm 0.06) \times 10^{-6}$ Å$^{-2}$. In contrast, the nSLD for CoFeB remains unchanged at $(5.71 \pm 0.15) \times 10^{-6}$ Å$^{-2}$.

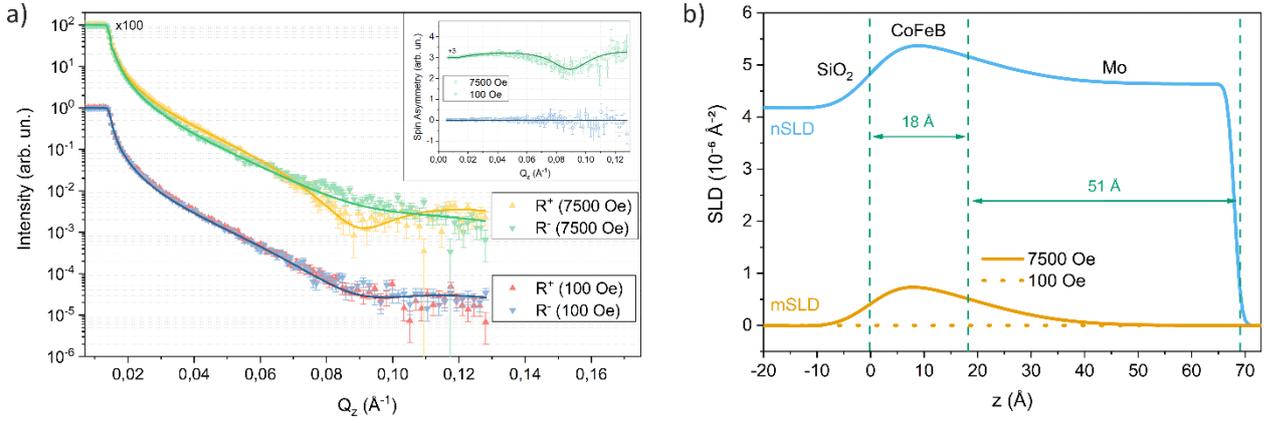

*Fig. 3. a) Reflectometry curves for the CoFeB / Mo (51 Å) bilayer structure, measured at 100 and 7500 Oe. The inset shows the SA curves for both cases (the measured data are represented by dotted lines and the fitted models are represented by solid lines). b) nSLD and mSLD profiles for the CoFeB / Mo (51 Å) film, measured at a field of 7500 Oe.*

Such behavior indicates the formation of PMA in the CoFeB / Mo structure. Earlier, PMA was studied for NiFe-based films with a Ru capping layer grown on SiO$_2$, and it was shown that the anisotropy at the Ox / FM interface exceeds the anisotropy at the FM / HM interface [35]. For the CoFeB / Pt (2 nm) bilayer grown on Si / SiO$_2$, PMA appears only after annealing, probably due to the formation of the Fe–O bonds [36]. Our results demonstrate that PMA is formed just after the deposition of a molybdenum layer on top of the ferromagnet. Thus, in the SiO$_2$ / CoFeB / Mo structure, the FM / HM interface is decisive in forming perpendicular magnetization. Furthermore, we show that PMA in the CoFeB/Mo system is achieved without annealing of the bilayer structure. For the first time, we demonstrate the possibility of forming PMA exclusively at the interface between CoFeB and Mo without external thermal treatment of the structure. This result is important because most studies obtain magnetic anisotropy only after the annealing of the structures. This finding may open new horizons for the design and application of spintronic devices.

## 2. PMA in CoFeB / Mo (thin HM layer)

Now, we tried to grow a thinner molybdenum top layer to find the critical value of its thickness, which is still sufficient to form PMA. Similar to the experiment described in the previous paragraph, we first grew a thin CoFeB film and characterized its structural and magnetic properties by PNR measurements at a magnetic field of 100 Oe. The film thickness was $18 \pm 2$ Å and mSLD was $(1.29 \pm 0.13) \times 10^{-6}$ Å$^{-2}$, corresponding to $450 \pm 50$ emu/cm$^3$.

Subsequently, a 5.0 ± 1.2 Å thick molybdenum layer was deposited on top of the CoFeB film. The reflectivity curves for the CoFeB / Mo (5 Å) bilayer structure (Figure 4a) demonstrate similar results to those obtained for the structure with a 51 Å molybdenum top layer. At 100 Oe, the structure does not show any splitting between the $R^+$ and $R^-$ curves, indicating an absence of any in-plane magnetic moment. At 7500 Oe, the splitting appears, indicating the reorientation of magnetic moments into the film plane. We attribute this behavior to the formation of PMA in CoFeB / Mo, despite the molybdenum being only 5 Å thick. From the mSLD and nSLD profiles measured at 7500 Oe (Figure 4b), we determined that the mSLD value in this case was (1.03 ± 0.05)x10$^{-6}$ Å$^{-2}$, corresponding to 363 ± 18 emu/cm$^3$. The nSLD for the CoFeB and Mo layers were (5.71 ± 0.15)x10$^{-6}$ and (5.0 ± 0.4)x10$^{-6}$ Å$^{-2}$, respectively.

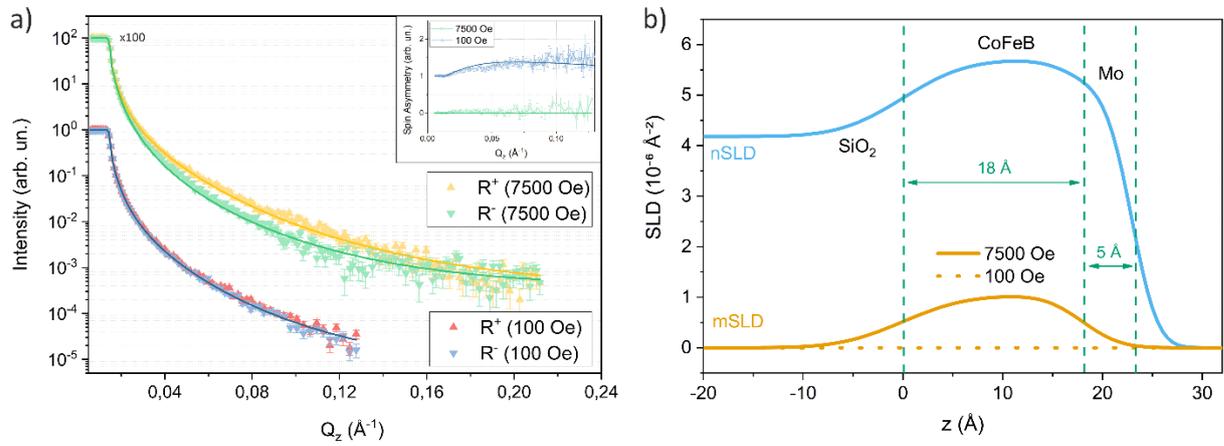

*Fig. 4. a) Reflectometry curves for the CoFeB / Mo (5 Å) bilayer structure, measured at 100 and 7500 Oe. The inset shows the SA curves for both cases (the measured data are represented by dotted lines and the fitted models are represented by solid lines). (b) nSLD and mSLD profiles for the CoFeB / Mo (5 Å) film, measured at a field of 7500 Oe.*

Previously, such thin layers of heavy metals were used only as 'dusting layers' to enhance PMA efficiency in structures based on CoFeB / MgO [18,22]. These layers, due to the high thermal stability of Mo and W, allowed the structures to withstand higher annealing temperatures, improving magnetic properties [10,14]. However, our work demonstrates that the 5 Å thick Mo film already acts as a functional layer for the formation of PMA and plays a crucial role in it. The use of such ultrathin metallic films can lead to breakthroughs and discoveries in the field of spintronics.

### 3. Silicon 'protective' layer

To protect the CoFeB / Mo (5 Å) sample with PMA, we deposited a ~50 Å thick Si film on top of it. To check the protective capability of such a layer, we exposed the sample to the air and then studied the magnetic properties of the resulting CoFeB / Mo / Si structure by PNR at a magnetic field of 7500 Oe (Figure 5a). We expected that the silicon layer would prevent the oxidation of the metal film while maintaining the unique magnetic characteristics of the structure. However, our data demonstrate the absence of splitting between the R⁺ and R⁻ curves even at high magnetic fields. The nSLD and mSLD profiles are shown in Figure 5b.

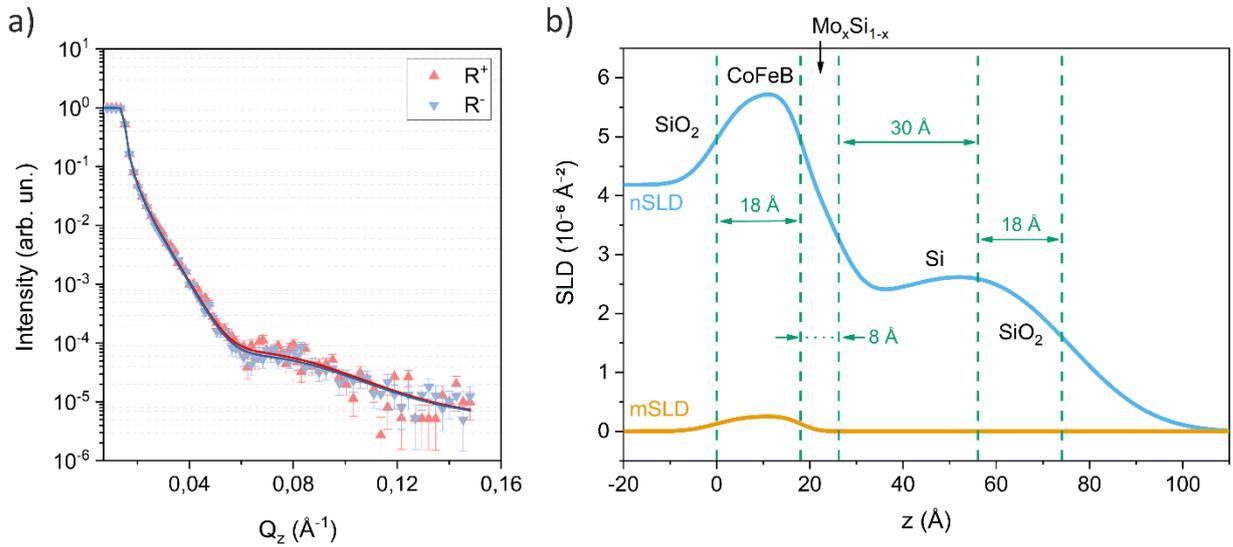

*Fig. 5. a) Reflectometry curves for the CoFeB / Mo (5 Å) / Si (SiO$_2$) film, measured at 7500 Oe (the measured data are represented by dotted lines and the fitted models are represented by solid lines). b) nSLD and mSLD profiles for the CoFeB / Mo (5 Å) / Si (SiO$_2$) trilayer structures, measured at a field of 7500 Oe.*

Neutron reflectometry data indicate that the top silicon layer is oxidized to a depth of approximately 18.0 ± 1.5 Å, which is consistent with the thickness of native oxide [38]. The thickness of a pure silicon layer is 33 ± 2 Å. The nSLD values for the Si and oxidized SiO$_2$ layers are (1.95 ± 0.44)x10⁻⁶ and (3.47 ± 0.71)x10⁻⁶ Å⁻², respectively. The molybdenum layer was fitted as a layer with silicon mixing, Mo$_x$Si$_{1-x}$ (x ~ 0.30–0.35), with an nSLD of (3.84 ± 0.41)x10⁻⁶ Å⁻². The thickness of mixing layer was 8.0 ± 1.3 Å. The layer of molybdenum is very thin and non-continuous, so Si can enter its pores during the growth [39]. Moreover, Si can diffuse into Mo even at room temperature [40]. The formation of such mixing layer might be responsible for disrupting the spin-orbit coupling between Mo and CoFeB, leading to the complete suppression of perpendicular magnetic anisotropy. Moreover, in the final structure, we observed neither PMA nor ferromagnetism in general assuming some degradation of the CoFeB layer as well.

*Conclusion*

In this work, we demonstrated the unique research capabilities of the pulsed laser deposition system combined with in-situ neutron reflectometry measurements for the investigation of magnetic properties of thin-film structures. This combination allowed us to examine in detail the mechanisms of perpendicular magnetic anisotropy formation in ultrathin metallic films. CoFeB films with thicknesses ranging from 11 to 18 Å were deposited on silicon substrates with a thermal oxide layer (Si / $SiO_2$). These single-layer films had an in-plane orientation of the magnetic moments. After the growth of a 53 Å Mo layer on top of the CoFeB film, perpendicular magnetic anisotropy was formed. PMA was observed in the sample without any thermal annealing of the structure. We demonstrated that thick molybdenum layers are not necessary for PMA formation in CoFeB films – a Mo layer of just 5 Å is enough for this purpose. Such ultrathin Mo film becomes not just a 'dusting layer' but a fully functional layer crucial for the formation of PMA. Finally, we showed that a capping Si layer of approximately 50 Å disrupted the magnetic properties of the CoFeB / Mo structure. The possible reason is the formation of mixing layer on the Mo-Si interface, which breaks the spin-orbit coupling between the CoFeB and Mo layers. Thus, our work presents new and unique results in the study of perpendicular magnetic anisotropy, achieved through in-situ neutron reflectometry measurements.

*Acknowledgements*

This work was prepared with support from the Ministry of Science and Higher Education of the Russian Federation (project FZWM-2024-0011).

The authors express their gratitude to the ILL, France, Grenoble, for providing beamtime on Super ADAM [41, 42].

*References*

[1] Y. Wang, X. Nie, J. Song, C. Wang, F. Yang, Y. Chi, X. Yang, Y. Shen, C. Xu, First-Principles Study of Origin of Perpendicular Magnetic Anisotropy in Mgo|Cofeb|Ta Structures, SSRN Journal (2022). https://doi.org/10.2139/ssrn.4302358.

[2] D. Lordan, G. Wei, P. McCloskey, C. O'Mathuna, A. Masood, Origin of perpendicular magnetic anisotropy in amorphous thin films, Sci Rep 11 (2021) 3734. https://doi.org/10.1038/s41598-020-78950-7.

[3] F. Hellman, A. Hoffmann, Y. Tserkovnyak, G.S.D. Beach, E.E. Fullerton, C. Leighton, A.H. MacDonald, D.C. Ralph, D.A. Arena, H.A. Dürr, P. Fischer, J. Grollier, J.P. Heremans, T. Jungwirth, A.V. Kimel, B. Koopmans, I.N. Krivorotov, S.J. May, A.K. Petford-Long, J.M. Rondinelli, N. Samarth, I.K. Schuller, A.N. Slavin, M.D. Stiles, O.

Tchernyshyov, A. Thiaville, B.L. Zink, Interface-induced phenomena in magnetism, Rev. Mod. Phys. 89 (2017) 025006. https://doi.org/10.1103/RevModPhys.89.025006.

[4] R. Sbiaa, H. Meng, S.N. Piramanayagam, Materials with perpendicular magnetic anisotropy for magnetic random access memory, Physica Rapid Research Ltrs 5 (2011) 413–419. https://doi.org/10.1002/pssr.201105420.

[5] B. Tudu, A. Tiwari, Recent Developments in Perpendicular Magnetic Anisotropy Thin Films for Data Storage Applications, Vacuum 146 (2017) 329–341. https://doi.org/10.1016/j.vacuum.2017.01.031.

[6] S. Peng, D. Zhu, J. Zhou, B. Zhang, A. Cao, M. Wang, W. Cai, K. Cao, W. Zhao, Modulation of Heavy Metal/Ferromagnetic Metal Interface for High-Performance Spintronic Devices, Adv Elect Materials 5 (2019) 1900134. https://doi.org/10.1002/aelm.201900134.

[7] S. Ikeda, K. Miura, H. Yamamoto, K. Mizunuma, H.D. Gan, M. Endo, S. Kanai, J. Hayakawa, F. Matsukura, H. Ohno, A perpendicular-anisotropy CoFeB–MgO magnetic tunnel junction, Nature Mater 9 (2010) 721–724. https://doi.org/10.1038/nmat2804.

[8] W.-G. Wang, M. Li, S. Hageman, C.L. Chien, Electric-field-assisted switching in magnetic tunnel junctions, Nature Mater 11 (2012) 64–68. https://doi.org/10.1038/nmat3171.

[9] Y. Liu, K.-G. Zhu, H.-C. Zhong, Z.-Y. Zhu, T. Yu, S.-D. Ma, Effect of Mo capping layers thickness on the perpendicular magnetic anisotropy in MgO/CoFeB based top magnetic tunnel junction structure, Chinese Phys. B 25 (2016) 117805. https://doi.org/10.1088/1674-1056/25/11/117805.

[10] T. Liu, Y. Zhang, J.W. Cai, H.Y. Pan, Thermally robust Mo/CoFeB/MgO trilayers with strong perpendicular magnetic anisotropy, Sci Rep 4 (2014) 5895. https://doi.org/10.1038/srep05895.

[11] M. Li, J. Lu, M. Akyol, X. Chen, H. Shi, G. Han, T. Shi, G. Yu, A. Ekicibil, N. Kioussis, P.V. Ong, P.K. Amiri, K.L. Wang, The impact of Hf layer thickness on the perpendicular magnetic anisotropy in Hf/CoFeB/MgO/Ta films, Journal of Alloys and Compounds 694 (2017) 76–81. https://doi.org/10.1016/j.jallcom.2016.09.309.

[12] T. Liu, J.W. Cai, L. Sun, Large enhanced perpendicular magnetic anisotropy in CoFeB/MgO system with the typical Ta buffer replaced by an Hf layer, AIP Advances 2 (2012) 032151. https://doi.org/10.1063/1.4748337.

[13] G.-G. An, J.-B. Lee, S.-M. Yang, J.-H. Kim, W.-S. Chung, J.-P. Hong, Highly stable perpendicular magnetic anisotropies of CoFeB/MgO frames employing W buffer and capping layers, Acta Materialia 87 (2015) 259–265. https://doi.org/10.1016/j.actamat.2015.01.022.

[14] Y.Q. Guo, H. Bai, Q.R. Cui, L.M. Wang, Y.C. Zhao, X.Z. Zhan, T. Zhu, H.X. Yang, Y. Gao, C.Q. Hu, S.P. Shen, C.L. He, S.G. Wang, High thermal stability of perpendicular magnetic anisotropy in the MgO/CoFeB/W thin films, Applied Surface Science 568 (2021) 150857. https://doi.org/10.1016/j.apsusc.2021.150857.

[15] D.C. Worledge, G. Hu, D.W. Abraham, J.Z. Sun, P.L. Trouilloud, J. Nowak, S. Brown, M.C. Gaidis, E.J. O'Sullivan, R.P. Robertazzi, Spin torque switching of perpendicular Ta|CoFeB|MgO-based magnetic tunnel junctions, Applied Physics Letters 98 (2011) 022501. https://doi.org/10.1063/1.3536482.

[16] P.F. Carcia, Perpendicular magnetic anisotropy in Pd/Co and Pt/Co thin-film layered structures, Journal of Applied Physics 63 (1988) 5066–5073. https://doi.org/10.1063/1.340404.

[17] W. Du, M. Liu, G. Wang, H. Su, B. Liu, H. Meng, X. Tang, Stack structure and CoFeB composition dependence of perpendicular magnetic anisotropy employing Pt heavy metal layer, Journal of Alloys and Compounds 928 (2022) 167205. https://doi.org/10.1016/j.jallcom.2022.167205.

[18] J.M. Iwata-Harms, G. Jan, S. Serrano-Guisan, L. Thomas, H. Liu, J. Zhu, Y.-J. Lee, S. Le, R.-Y. Tong, S. Patel, V. Sundar, D. Shen, Y. Yang, R. He, J. Haq, Z. Teng, V. Lam, P. Liu, Y.-J. Wang, T. Zhong, H. Fukuzawa, P.-K. Wang, Ultrathin perpendicular magnetic anisotropy CoFeB free layers for highly efficient, high speed writing in spin-transfer-torque magnetic random access memory, Sci Rep 9 (2019) 19407. https://doi.org/10.1038/s41598-019-54466-7.

[19] J. Hayakawa, S. Ikeda, Y.M. Lee, R. Sasaki, T. Meguro, F. Matsukura, H. Takahashi, H. Ohno, Current-Driven Magnetization Switching in CoFeB/MgO/CoFeB Magnetic Tunnel Junctions, Jpn. J. Appl. Phys. 44 (2005) L1267. https://doi.org/10.1143/JJAP.44.L1267.

[20] H. Sato, E.C.I. Enobio, M. Yamanouchi, S. Ikeda, S. Fukami, S. Kanai, F. Matsukura, H. Ohno, Properties of magnetic tunnel junctions with a MgO/CoFeB/Ta/CoFeB/MgO recording structure down to junction diameter of 11 nm, Applied Physics Letters 105 (2014) 062403. https://doi.org/10.1063/1.4892924.

[21] L. Saravanan, N.K. Gupta, L. Pandey, I.P. Kokila, H.A. Therese, S. Chaudhary, Observation of uniaxial magnetic anisotropy and out-of-plane coercivity in W/Co20Fe60B20/W structures with high thermal stability, Journal of Alloys and Compounds 895 (2022) 162600. https://doi.org/10.1016/j.jallcom.2021.162600.

[22] T. Yamamoto, T. Ichinose, J. Uzuhashi, T. Nozaki, T. Ohkubo, K. Yakushiji, S. Tamaru, S. Yuasa, Large Tunneling Magnetoresistance in Perpendicularly Magnetized Magnetic Tunnel Junctions Using Co 75 Mn 25 / Mo / Co 20 Fe 60 B 20 Multilayers, Phys. Rev.

Applied 19 (2023) 024020. https://doi.org/10.1103/PhysRevApplied.19.024020.

[23] Q. Shao, P. Li, L. Liu, H. Yang, S. Fukami, A. Razavi, H. Wu, K. Wang, F. Freimuth, Y. Mokrousov, M.D. Stiles, S. Emori, A. Hoffmann, J. Akerman, K. Roy, J.-P. Wang, S.-H. Yang, K. Garello, W. Zhang, Roadmap of Spin–Orbit Torques, IEEE Trans. Magn. 57 (2021) 1–39. https://doi.org/10.1109/TMAG.2021.3078583.

[24] J. Sinova, S.O. Valenzuela, J. Wunderlich, C.H. Back, T. Jungwirth, Spin Hall effects, Rev. Mod. Phys. 87 (2015) 1213–1260. https://doi.org/10.1103/RevModPhys.87.1213.

[25] A. Hoffmann, Spin Hall Effects in Metals, IEEE Trans. Magn. 49 (2013) 5172–5193. https://doi.org/10.1109/TMAG.2013.2262947.

[26] L. Zhu, D.C. Ralph, R.A. Buhrman, Spin-Orbit Torques in Heavy-Metal–Ferromagnet Bilayers with Varying Strengths of Interfacial Spin-Orbit Coupling, Phys. Rev. Lett. 122 (2019) 077201. https://doi.org/10.1103/PhysRevLett.122.077201.

[27] T. Zhu, Y. Yang, R.C. Yu, H. Ambaye, V. Lauter, J.Q. Xiao, The study of perpendicular magnetic anisotropy in CoFeB sandwiched by MgO and tantalum layers using polarized neutron reflectometry, Applied Physics Letters 100 (2012) 202406. https://doi.org/10.1063/1.4718423.

[28] A. Vorobiev, A. Devishvilli, G. Palsson, H. Rundlöf, N. Johansson, A. Olsson, A. Dennison, M. Wollf, B. Giroud, O. Aguettaz, B. Hjörvarsson, Recent upgrade of the polarized neutron reflectometer Super ADAM, Neutron News 26 (2015) 25–26. https://doi.org/10.1080/10448632.2015.1057054.

[29] G.P. Felcher, S.G.E. Te Velthuis, A. Rühm, W. Donner, Polarized neutron reflectometry: recent developments and perspectives, Physica B: Condensed Matter 297 (2001) 87–93. https://doi.org/10.1016/S0921-4526(00)00821-8.

[30] J.F. Ankner, G.P. Felcher, Polarized-neutron reflectometry, Journal of Magnetism and Magnetic Materials 200 (1999) 741–754. https://doi.org/10.1016/S0304-8853(99)00392-3.

[31] A. Glavic, M. Björck, *GenX 3* : the latest generation of an established tool, J Appl Crystallogr 55 (2022) 1063–1071. https://doi.org/10.1107/S1600576722006653.

[32] Y. Zhu, Modern Techniques for Characterizing Magnetic Materials, 1st ed, Springer, New York, NY, 2005.

[33] A.J. Dianoux, G.H. Lander, Institut Laue-Langevin, eds., Neutron data booklet, 2nd ed, Old City, Philadelphia, PA, 2003.

[34] J.A.C. Bland, R.D. Bateson, B. Heinrich, Z. Celinski, H.J. Lauter, Spin polarised neutron reflection studies of ultrathin magnetic films, Journal of Magnetism and Magnetic Materials 104–107 (1992) 1909–1912. https://doi.org/10.1016/0304-8853(92)91600-X.

[35] J. Beik Mohammadi, G. Mankey, C.K.A. Mewes, T. Mewes, Strong interfacial

[35] perpendicular anisotropy and interfacial damping in Ni0.8Fe0.2 films adjacent to Ru and SiO2, Journal of Applied Physics 125 (2019) 023901. https://doi.org/10.1063/1.5052334.

[36] W.C. Tsai, S.C. Liao, H.C. Hou, C.T. Yen, Y.H. Wang, H.M. Tsai, F.H. Chang, H.J. Lin, C.-H. Lai, Investigation of perpendicular magnetic anisotropy of CoFeB by x-ray magnetic circular dichroism, Applied Physics Letters 100 (2012) 172414. https://doi.org/10.1063/1.4707380.

[37] H. Cheng, J. Chen, S. Peng, B. Zhang, Z. Wang, D. Zhu, K. Shi, S. Eimer, X. Wang, Z. Guo, Y. Xu, D. Xiong, K. Cao, W. Zhao, Giant Perpendicular Magnetic Anisotropy in Mo-Based Double-Interface Free Layer Structure for Advanced Magnetic Tunnel Junctions, Adv Elect Materials 6 (2020) 2000271. https://doi.org/10.1002/aelm.202000271.

[38] N.F. Mott, S. Rigo, F. Rochet, A.M. Stoneham, Oxidation of silicon, Philosophical Magazine B 60 (1989) 189–212. https://doi.org/10.1080/13642818908211190.

[39] P. Ohresser, J. Shen, J. Barthel, M. Zheng, Ch.V. Mohan, M. Klaua, J. Kirschner, Growth, structure, and magnetism of fcc Fe ultrathin films on Cu(111) by pulsed laser deposition, Phys. Rev. B 59 (1999) 3696–3706. https://doi.org/10.1103/PhysRevB.59.3696.

[40] D. Bhattacharyya, A.K. Poswal, M. Senthilkumar, P.V. Satyam, A.K. Balamurugan, A.K. Tyagi, N.C. Das, Surface roughness and interface diffusion studies on thin Mo and W films and Mo/Si and W/Si interfaces, Applied Surface Science 214 (2003) 259–271. https://doi.org/10.1016/S0169-4332(03)00360-X.

[41] VOROBIEV Alexei; Olga Dikaya; DOLGOBORODOV Artur; GOIKHMAN Alexander; GRUNIN Aleksei; KIRICHUK Grigorii and K.Maksimova. (2024). In situ growth and investigation of FeCoB thin films for energy storage applications. Institut Laue-Langevin (ILL) doi:10.5291/ILL-DATA.CRG-3091

[42] GOIKHMAN Alexander; DOLGOBORODOV Artur; GRUNIN Aleksei; KIRICHUK Grigorii; K.Maksimova and VOROBIEV Alexei. (2024). FeCoB/Mo and FeCoBo/Mo/NiO multilayers magnetic properties investigation upon In-situ growth and and in-operando heating by PNR. Institut Laue-Langevin (ILL) doi:10.5291/ILL-DATA.CRG-3164

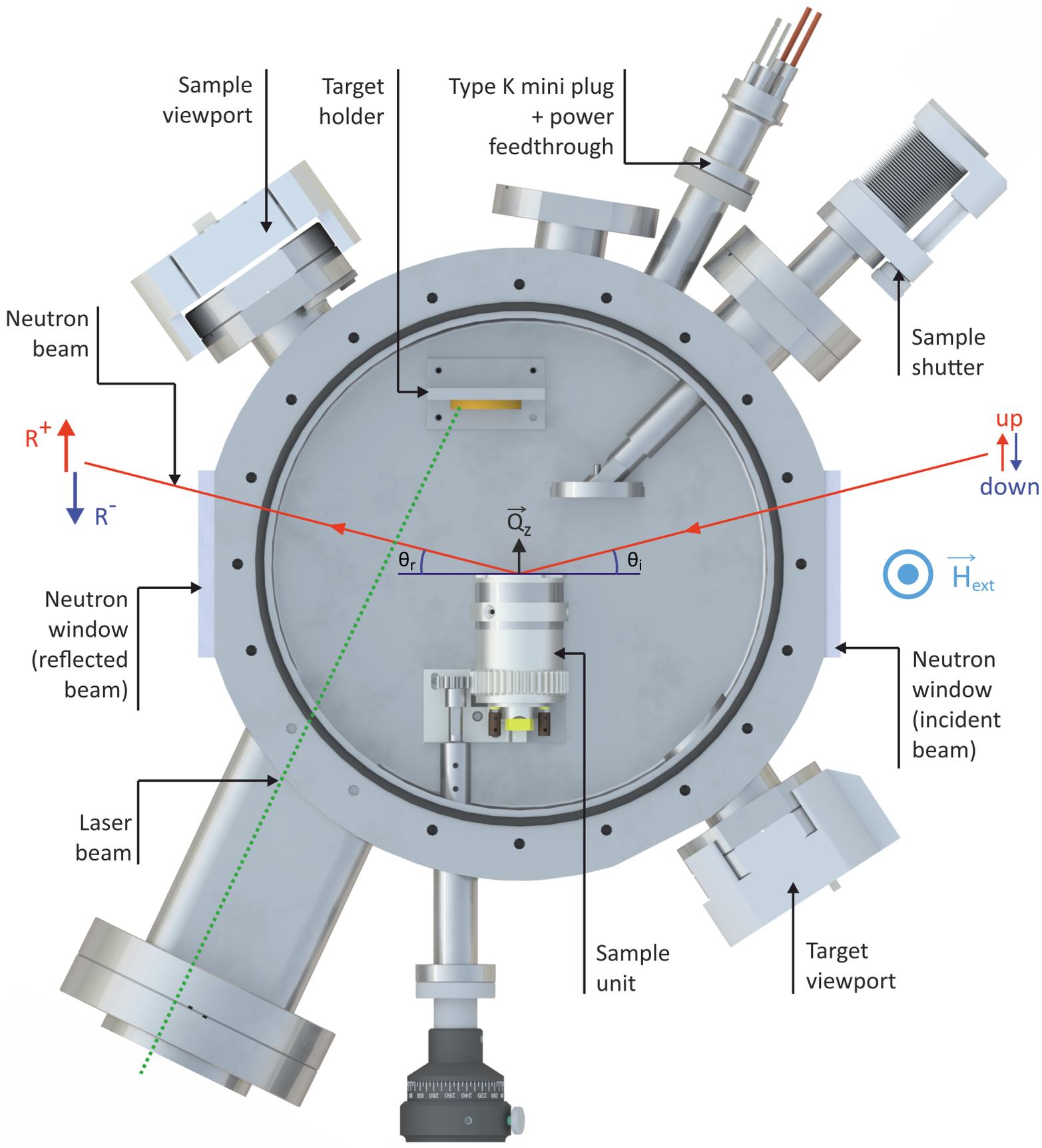

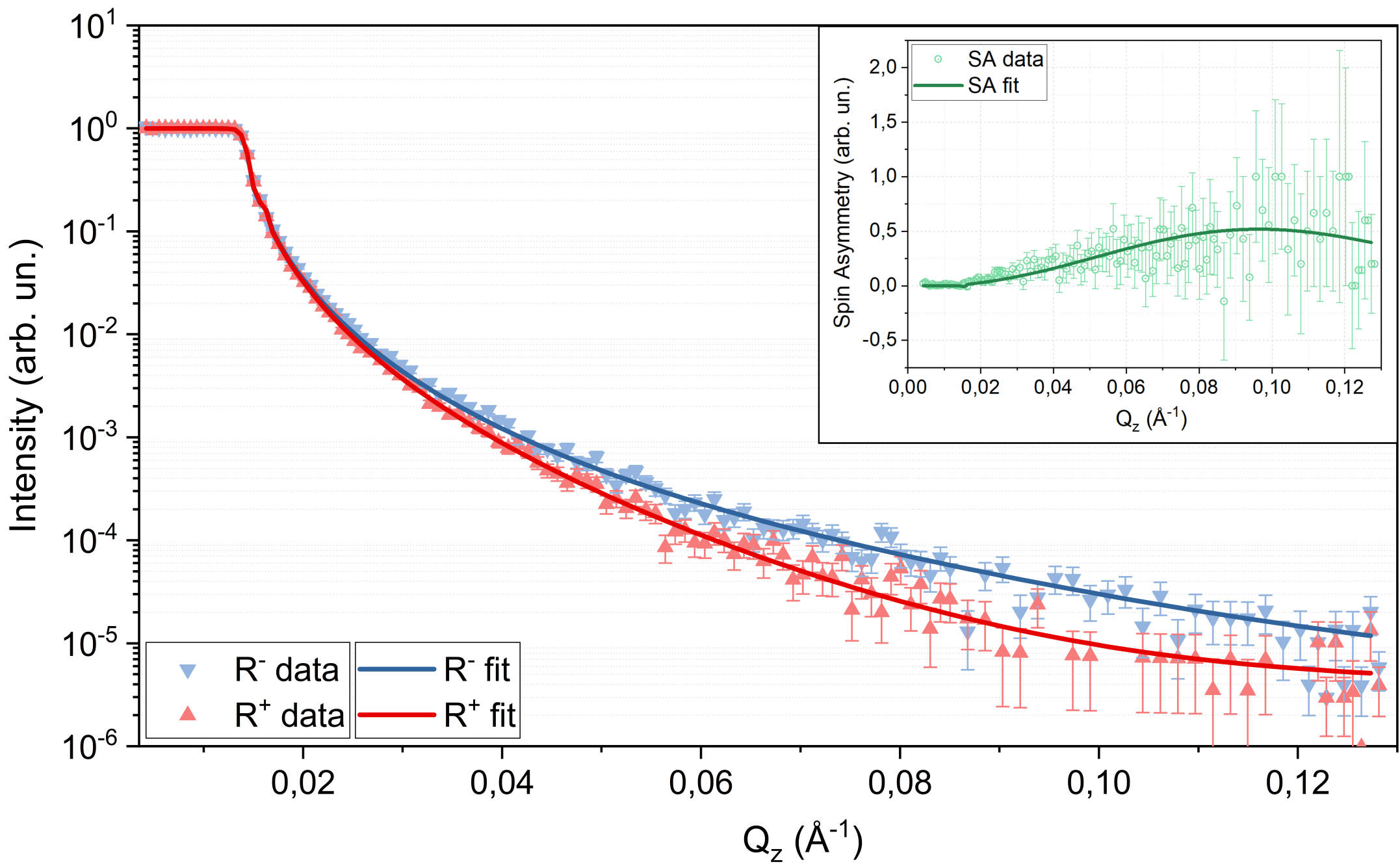

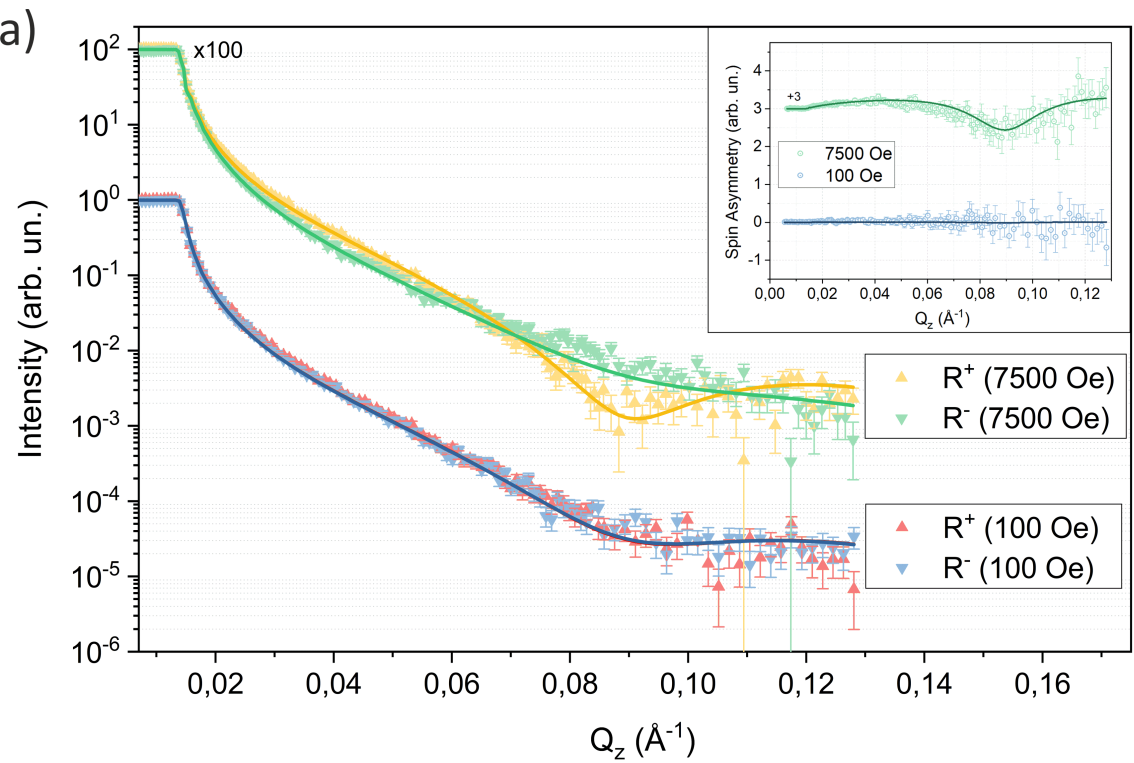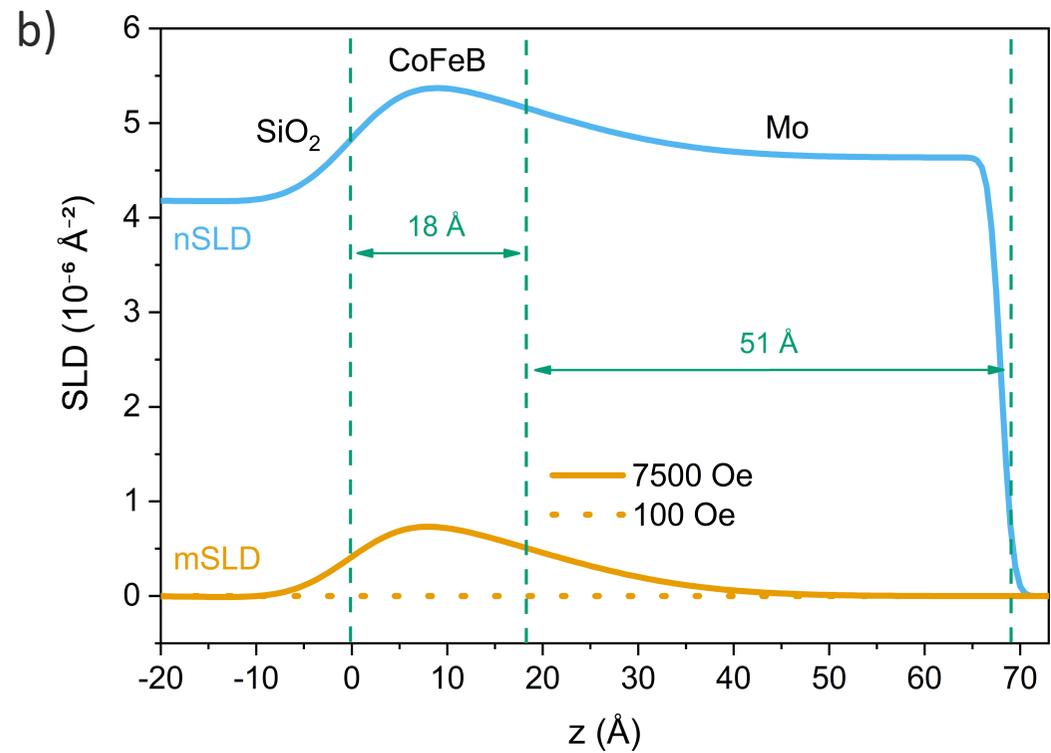

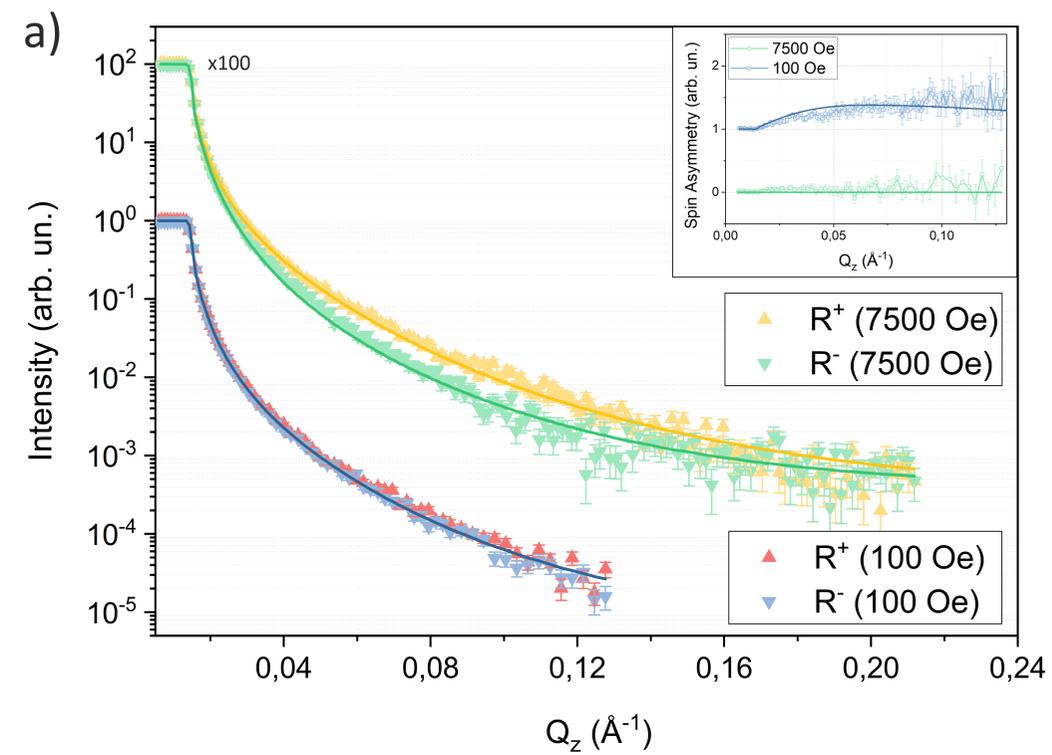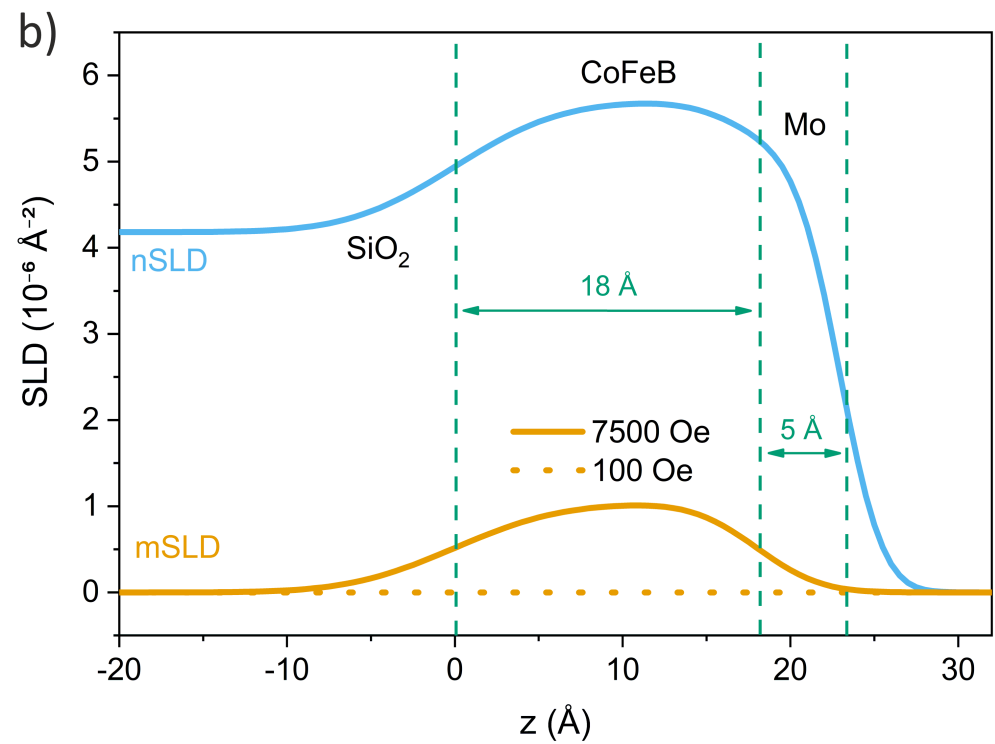

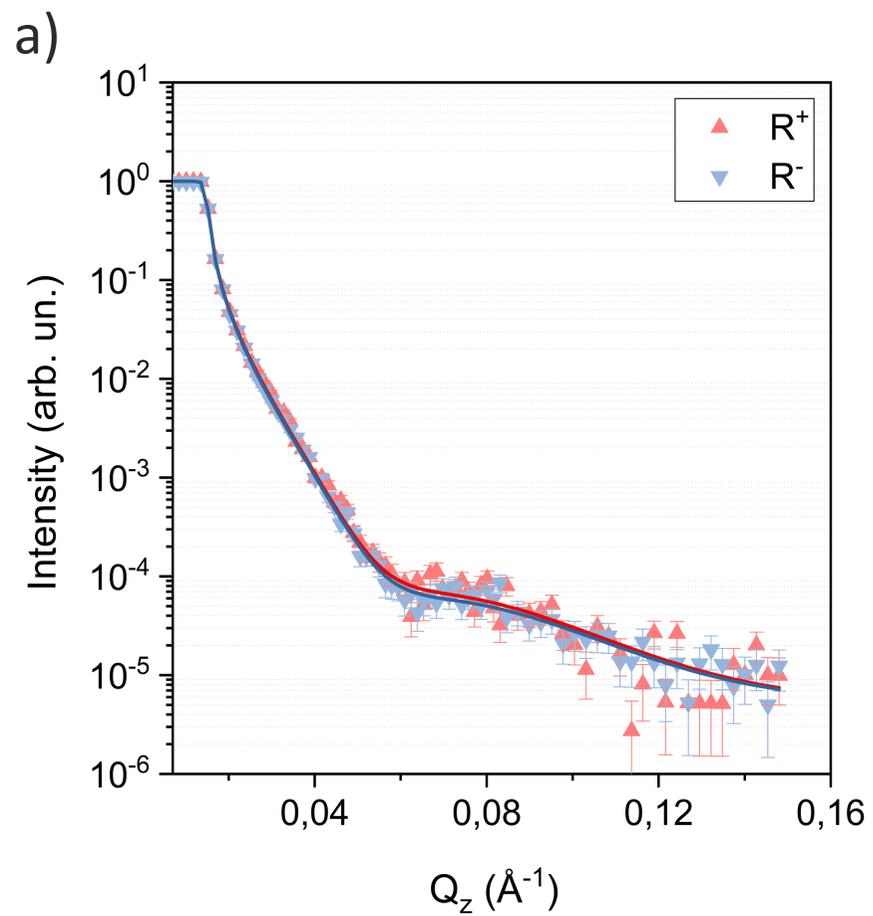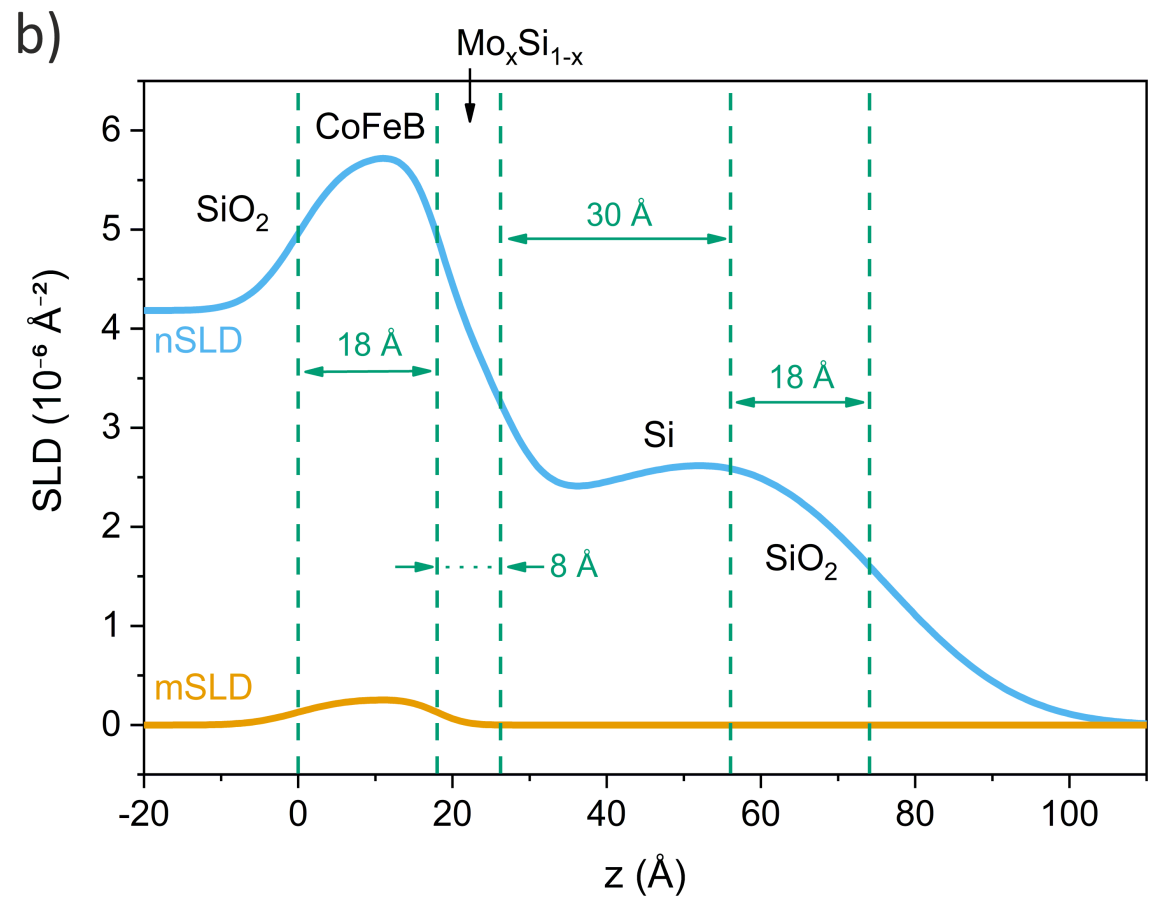